\documentclass[aps,pre,superscriptaddress,a4paper,preprint,showpacs]{revtex4-1}
\pdfoutput=1
\usepackage{amsmath}
\usepackage{amsfonts}
\usepackage{amssymb}
\usepackage[nooneline]{subfigure}
\usepackage{graphicx}
\usepackage{caption}
\usepackage{blindtext}
\usepackage{epstopdf}

\begin{document}

\title{Simulated annealing approach to vascular structure with application to the coronary arteries}

\author{Jonathan Keelan}
\affiliation{Department of Physical Sciences, The Open University, Milton Keynes, UK, MK7 6AA}

\author{Emma M. L. Chung}
\affiliation{Department of Cardiovascular Sciences, University of Leicester, UK, LE1 5WW}

\author{James P. Hague}
\email[Author to whom correspondence should be sent, e-mail:]{Jim.Hague@open.ac.uk}
\affiliation{Department of Physical Sciences, The Open University, Milton Keynes, UK, MK7 6AA}

\
\begin{abstract}
Does the complex processes of angiogenesis during organism development ultimately lead to a near optimal coronary vasculature in the organs of adult mammals? We examine this hypothesis using a powerful and universal method, built on physical and physiological principles, for the determination of globally energetically optimal arterial trees. The method is based on simulated annealing, and can be used to examine arteries in hollow organs with arbitrary tissue geometries. We demonstrate that the approach can generate {\it in-silico} vasculatures which closely match porcine anatomical data for the coronary arteries on all length scales, and that the optimised arterial trees improve systematically as computational time increases. The method presented
here is general, and could in principle be used to examine the arteries of other organs. Potential applications include improvement of medical imaging analysis and the design of vascular trees for artificial
organs.
\end{abstract}

\date{\today}






\maketitle

\section{Introduction}

 Arterial trees are vital for the efficient transport of oxygen and nutrients to tissue. Their anatomy has been studied for many centuries through the dissection of cadavers, inspection of corrosion casts, medical imaging techniques, and computational models.
It has been determined that individual
arterial bifurcations follow optimality principles that lower
metabolic demand locally \cite{ZamirBook,Frame,Murray1926a,Rossitti1993,Kassab1993,Cassot2010}, as demonstrated by the scaling laws
followed by arterial trees \cite{Huo2012,Huo2009,Kassab2006,Bengtsson2003,West1997}.
More recently, there has been a high level of interest in models that mimic arterial growth (angiogenesis) using physical and physiological principles to simulate vascular anatomy.
These models are created based on local optimisation principles, where the anatomy of each branch in the arterial tree is governed by a compromise between maximising fluid dynamical efficiency and minimising the quantity of blood required. However, models of coronary
vasculature, based on local optimisation are not able to explain if the organisation of major arteries is the result of fluid dynamical optimisation across the `whole organ' \cite{Karch1999,Karch2000,Kaimovitz2005,Schreiner1993,Kaimovitz2010}.

The relationship between the radii of vessels in individual bifurcations is well categorised by the equation:
$r_{p}^\gamma = r_{d_1}^\gamma +
r_{d_2}^\gamma$ where $r_p$ is the radius of the parent artery, $r_{d_{1,2}}$ those of
the daughter arteries, and $\gamma$ is the bifurcation exponent\cite{Mayrovitz1983,Rossitti1993,Nakamura2014}, which has the value $3.0$ in Murray's formulation. This relation arises from the combination of the continuity equation and the diameter-flow rate relation\cite{Kassab2006}. Several current  methods for in-silico growth of vascular trees into a simulated tissue substrate aim to optimise the local properties of individual bifurcations \cite{Schreiner1993,Karch2000a}. A  procedure, known as constrained constructive optimisation (CCO) starts by inserting a single artery into the tissue. A new vessel with a position chosen at random is then connected to the original artery and the link point is moved such that the energy of the arteries is minimised. New arteries are then iteratively added and optimised until a predetermined number of terminal sites have been added.
The overall result is that CCO and similar methods create trees whose structure is predetermined by the order in which new arteries are added: if the order is changed, the final tree structure also changes. Morphologically, CCO reproduces a reasonable distribution of vessel sizes due to the application of Murray's law, but creates arterial branches that are more symmetrical than those found in nature (especially for the largest arteries) \cite{Schreiner1997} and significant extensions are required to generate vessels in hollow organs \cite{Schreiner2006}. The variations in the structure and positions of larger arteries in CCO generated trees are problematic, since organs such as the heart exhibit only small differences in large artery structure over a population (aside from rare abnormalities).
An alternative method known as global constructive optimisation (GCO) attempts to overcome the problems of CCO by including a multiscale pruning update that is global in the sense that it acts simultaneously on a significant subset of the tree, but otherwise only includes updates allowing local modifications to the topology of the tree. As such, GCO is limited to sampling a subset of the allowed topologies of the arterial trees \cite{Georg2010}, so while it is expected to offer improvements over CCO, it carries no guarantee of reaching the global minimum. Due to the use of local downhill searches, GCO also has similar issues with hollow organs. To obtain a universal optimisation technique to compute in-silico arterial trees for arbitrary tissue structures, a different approach is needed.

Another method for the generation of large scale arterial trees uses extensive morphological databases\cite{Kaimovitz2005, Kassab1997}. These trees contain far more vessels than is feasible to generate with techniques such as CCO, as the topology of the tree is taken from experimental data.  However, since detailed morphological databases do not exist for the vast majority of organs, the use of these techniques is impossible in the general case. Morphologically generated models provide trees suitable for large scale fluid dynamical studies and organ phantoms\cite{Fung2011}. They achieve this by reproducing experimental data in a computationally accessible form. As such they have no predictive powers that can contribute to the understanding of the origins of arterial tree structure.

A separate class of models exist for use in modeling the growth of tumorous vasculature and the process of vascular remodeling, which involve the direct simulation of sprouting angiogenesis\cite{ref1,ref2,ref3,ref4}. These models seek to reproduce the rapid and dynamic process of tumor vascularisation, or the growth of vasculature in normal tissue as it grows. Using in silico simulation of sprouting angiogenesis to obtain the vascular structure in a fully grown organ would be extremely difficult, as full details of the distribution of tissue and oxygen demands would be needed for all stages of embryonic and childhood development. As organs such as the heart have very small levels of variation in vascular structure over the population, the structure itself is likely to be  caused by a process different to that of tumor vascularisation. We suggest that this process is an optimum seeking one, and that as such an optimisation procedure is required to accurately model it. We examine if, regardless of the complex processes that guide angiogenesis during growth, the final structure of the coronary vasculature in adult mammals is near optimal.

Development of a method which reaches a morphologically accurate solution based solely upon optimisation criteria would be useful in vascular research, allowing for the modeling of realistic vascular trees in organs lacking extensive morphological databases. The inability of CCO and extensions to find the global energy minimum, and the subsequent lack of consistent structure (particularly of the larger arteries), is problematic if organ specific vasculature is required. An approach capable of producing an arterial tree, which minimises pumping power and blood volume, whilst providing adequate blood flow to critical regions would be invaluable in this regard. This paper goes beyond previous work by introducing a far more flexible and universal method for generation of `whole organ' arterial trees, in any arbitrarily shaped tissue substrate, that obey both local and global optimisation criteria. To identify globally optimised arterial trees, we use a powerful computational technique known as Simulated Annealing (SA) \cite{kirkpatrick1983a}. Although SA is computationally expensive, correctly applied SA techniques have a key advantage of being mathematically and computationally proven to converge to a global energy minimum. To achieve this, our SA based approach has the potential to sample all possible arterial tree configurations, ranging from perfectly symmetric, intricately bifurcating structures, to asymmetric trees characterised by a single trunk vessel. This is achieved by allowing: (1) repositioning of bifurcations, and (2) swapping the parent vessels of bifurcations between different parts of the tree. By introducing these forms of plasticity to our models, the entire parameter space of the tree can be explored, allowing the method to identify the best possible arterial configuration for supplying a particular organ. Full details of this novel method can be found at the end of the article. As an example application we determine the near optimal configuration of arteries for supplying the heart and compare our computer generated coronary vasculature with morphological data from real coronary arteries. Specifically, we determine that the observed anatomy of the coronary arteries is similar to that expected from near global minimisation of total energy expenditure, and validate the approach against porcine data, finding a very high level of agreement with morphological data.

\section{Method}
\label{sec:method}

The main purpose of any arterial tree is to maintain adequate blood perfusion with minimal total metabolic expense. The suitability of an arterial tree for this purpose is governed by two considerations:  (1) since blood is viscous, the power required to pump blood through the vasculature should be minimised, (2) as energy is required to generate and maintain blood, the volume of blood required should be minimised. Murray's law achieves this for individual bifurcations, but the optimal organisation of large numbers of connected bifurcations is far from obvious. The interplay between these competing concerns for thousands of arterial segments leads to a complex optimisation problem. Note that in the following, bifurcations will be referred to as nodes,
arteries will be referred to as `segments between nodes', and terminal
arterioles are referred to as `end nodes'.

\subsection{Metabolic cost to maintain blood volume}

The first component of the approach involves calculating the power needed to maintain the entire tree, which will be used as a value in the cost function. The power consumption of the tree can be split into two separate parts: the first is the metabolic cost of maintaining the blood volume and tissue associated with the tree, and the second is the power required to pump blood through the tree. The length and radius of each segment (vessel) $i$ of the tree must be known to calculate the volume. By assuming a fixed bifurcation exponent, the radii are determined by the topology and only vessel lengths rely on the geometrical arrangement. To calculate the cost, volume must be multiplied by a constant, $m_b$, corresponding to a physiologically reasonable metabolic demand of the same quantity of blood and vascular tissue \cite{Liu2007}. Thus the metabolic cost due to the volume of the tree will be given by:
\begin{equation}
C_v = m_b V_{\textit{tree}}
\end{equation}
where $m_b$ is taken to be 641.3 J s$^{-1}$ m$^{-3}$ and $V_{\textit{tree}}$ is the volume of the entire tree.

\subsection{Power cost to pump blood through vessels}

To calculate the power needed to pump blood through the entire tree, we must know the pressure and volumetric flows inside each segment (vessel) of the tree, which can be found by first assuming that Poiseuille's law,  $\Delta p = Q R$, is followed inside the segments, where $\Delta p$ is the pressure drop over the vessel, and $Q$ is the flow. The assumption that flow is laminar inside the vessels is justified provided that the typical length of a vessel is much larger than the radius, and that pulsatile flow effects are negligible. Vessels within the simulated trees have a typical length radius ratio of $10$, and while in the largest arteries of the tree pulsatile effects may still be present, these rapidly decay so that the vast majority lie within a non-pulsatile regime. We assume both Murray's law and that terminal node flows are constant to simplify calculation of the relevant fluid dynamical quantities: the only quantity which relies on the structure of the tree is the pressure. In a sense, the segments can be considered as connected set of resistors, with the resistance given by:
  \begin{equation}
R = \frac{8 \mu L}{\pi r^4},
\end{equation}
where $r$ is the radius of the vessel, $L$ its length and $\mu=3.6\times 10^{-3} \mathrm{Pa}\,\mathrm{s}$ the viscosity of blood. The pressures (and hence flows) for every node in the tree can then be found recursively. $W_{i}$, the power consumed by each segment $i$ is then calculated using:
\begin{equation}
W_i = Q_i^2 R_i,
\end{equation}
Summing over all segments in the tree, the total power required to maintain the proper flow through the tree is:
\begin{equation}
C_w = \sum_i^{N_{\textit{tot}}}W_i.
\end{equation}

\subsection{Ensuring tissue supply}

\begin{figure}
\includegraphics[width=0.45\textwidth]{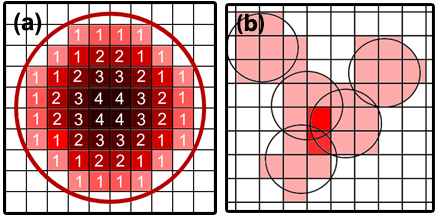}
\caption{(a) The distance map for a spherical surface. Voxels outside the surface have value 0, with those inside the surface contributing a value relating to their distance from the surface. (b) Each arteriole supplies a spherical region shown by the lightly shaded squares. Where there is significant overlap between two spheres, there is a penalty. Unsupplied voxels also incur a penalty in the cost function.}
\label{fig:distsupmap}
\end{figure}

The primary purpose of the vascular tree is to supply blood, thus it is important that terminal nodes are correctly dispersed inside the tissue. Initially, terminal nodes are randomly distributed inside the tissue, with each node having associated with it a sphere of influence for blood supply. The radius of this sphere is calculated using physiological values for the blood demand of the tissue. The density of myocardium is $\rho=1.06\times 10^{3}$kg m$^{-3}$ \cite{Vinnakota2004}, and the flow demand is $1.13$ ml min$^{-1}$g$^{-1}$ \cite{Uren1994} leading to a flow demand per m$^3$ of heart tissue of $q_{\rm required}=2\times 10^{-2}$s$^{-1}$. The total flow into the heart is $Q_0=4.16\times 10^{-6}$m$^{3}$s$^{-1}$ \cite{Hall2010}, which can be converted to total flow per node as $Q_N=Q_0/N$, where $N$ is the total number of arterioles (end nodes). The radius of the supply sphere is then calculated via $4\pi R_{\rm supply}^3 / 3 = Q_N/q_{\rm required}$. The sphere can be thought of as a microcirculatory `black box' \cite{Schreiner1993}, where the exact fluid dynamical details of the blood flow have been ignored. Spheres of blood supply associated with end nodes are stored in a voxel map (a voxel is a 3D generalisation of a pixel) of the tissue, where each terminal node adds exactly one to each voxel inside its sphere of supply (Fig \ref{fig:distsupmap}b). While blood demand is not constant within the myocardium at any single instance of time, the majority of fluctuations are high frequency oscillations which are assumed to be averaged out in the present model\cite{bfoscillations}.
The terminal nodes are then allowed to move inside the tissue, where after each move a new voxel supply map is calculated, and the overlap (each voxel supplied by more than 1 sphere
, or the dark red voxels in Fig \ref{fig:distsupmap}b
) is used as a value in the cost function of the simulated annealing algorithm. In addition, all voxels not being supplied are given a cost, so that the overall penalty associated with having both unsupplied and oversupplied voxels is chosen to be:
\begin{equation}
C_s = \sum_{\textbf{voxels}} s;\,
s = \left\{ \begin{array}{rl}
 10 &\mbox{ if $b = 0$} \\
  (b - 1)^2 &\mbox{ otherwise}
       \end{array} \right.
\end{equation}
where $b$ is the value of the supply at the voxel and the sum is performed over all the voxels comprising the tissue. In practice, this cost is set to be much larger than all other costs, since any unsupplied tissue would die. Therefore, the terminal nodes spread evenly through the tissue early in the optimisation. Other functions may be used, provided that the minimum in the function for each voxel occurs at $b=1$. The value $C_s$ then defines the fitness of the tree to supply blood, and the penalty for over supplying voxels forms a sort of self avoidance algorithm, where terminal nodes are encouraged to pack the tissue as densely as possible without overlapping. A benefit of this method is that it allows easy integration of medical imaging into the model, as well as providing an easy method for differentiating tissue with different blood supply demands.

\subsection{Exclusion of large vessels from tissue}

In order to create a realistic vascular tree, it must be possible to exclude some segments from penetrating the tissue. For instance, in the case of the heart, it would be unlikely to find a very large artery within the myocardium, and vessels may not penetrate the ventricles; rather, the larger arteries and arterioles lie on the surface of the heart, with only the smaller arterioles and capillaries being found inside the tissue. To mimic this structure, the approach makes use of a cut off radius $ R_{\rm{cutoff}} $, whereby segments with radius larger than $ R_{\rm{cutoff}} $ may not penetrate the tissue. In the calculations performed in this article, $R_{\rm{cutoff}}=0.01$ mm. To determine which segments with radius greater than $R_{\rm{cutoff}}$ have penetrated the tissue we first take a distance transform of the tissue surface for each tissue voxel (Fig \ref{fig:distsupmap}a. )
This provides a second voxel map of the tissue, distinct from the blood supply map, giving a measure of the distance of a point from the surface when it is inside the tissue (outside of the surface, the value is zero). For each segment satisfying the radius criteria, a list of voxels that its centre-line penetrates is generated \cite{Amanatides1987}, along with a value for the length element of the segment present inside that voxel. A cost is then calculated based upon the value of the distance transform at each of the voxels according to,
\begin{equation}
C_o = \pi r^2 (D_{ijk} \tilde{L}_{ijk})^6, \label{exclusioncost}
\end{equation}
where $i$, $j$ and $k$ are the cartesian voxel coordinates taken from the centre-line of the segment. $D_{ijk}$ is the value of the distance transform at that voxel coordinate. $\tilde{L}_{ijk}$ is the length of the segment spent inside the voxel. The sum is performed over all the voxels contained in the list calculated from the centre-line. This cost can then be used in the SA algorithm as a penalty that favours moving large segments out of the tissue.

\subsection{Pressure constraints}

In physiologically realistic trees, capillary networks should receive a constant pressure  $ P_{\textit{term}} $ to function correctly. A new cost can be devised to ensure this. A suitable candidate is,
\begin{equation}
C_p = \sum_i^{N_{\textit{term}}} (P_i - P_{\textit{term}})^2,
\end{equation}
where the sum is performed over all terminal nodes, and  $ P_i $ is the actual terminal node pressure. In practice, for trees which can be optimised on feasible time scales (i.e of a few thousand nodes), the pressure drop from root to end node is less than 1\% of the total pressure drop of a real arterial tree, with most of the pressure drop occurring over smaller arterioles than those considered here, so it is unnecessary to perform this calculation. When it becomes possible to grow larger trees, the pressure at the capillaries will need to be taken into consideration. This will add a significant computational cost.

\subsection{Total cost function}

We have now determined a form for all the relevant costs associated with an arbitrary tree configuration supplying arbitrary tissue shapes. We can therefore define a total cost which gives a numeric measure of the fitness of a given tree,
\begin{equation}
C_{\textit{T}} = A_{w,v}(C_w + C_v) + A_o C_o + A_p C_p + A_sC_s
\end{equation}
where $A_i$ indicates a weighting value which scales each relevant cost. There is no way to analytically determine what weights to use, and the selection of appropriate weights must found experimentally, however a few basic principles such as the having a very high weight for the blood supply cost and a low weight for the end node pressure cost can guide the process. In principle, $A_{s}$ should be infinite, since tissue without supply dies. In this work, we use $A_{w,v}=1 \times 10^{4}$, $A_p=0$, $A_s=1\times 10^{30}$ and $A_o=100$. In this way, $A_s$ and $A_o$ force the exclusion of vessels and uniform supply of tissue to act like constraints. While the exclusion cost should technically be infinite, as no arteries are found in the ventricles of living humans, it is advantageous to give it a large but finite value. This allows the optimisation procedure to identify gradients, giving it extra information and speeding up convergence. This size of the constant for $A_o$ may seem small in this regard, however its value $C_o$ is already raised to the power $6$ in Eq \ref{exclusioncost}. Since any scaling of the cost function does not effect the location of minima, we can absorb one of the weightings by scaling everything else. This allows a new cost function $\tilde{C}_T = \frac{C_T}{A_{w,v}}$ to be defined.

\subsection{Simulated annealing}

\begin{figure}
\includegraphics[width=0.45\textwidth]{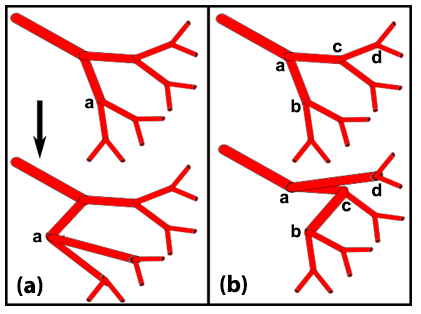}
\caption{Tree modification updates}
\label{fig:samoves}
\end{figure}

To select the fittest, most optimised trees, we use a powerful technique for optimisation problems known as simulated annealing (SA) \cite{Henderson2003,Kirkpatrick1983}.
The
primary difference between SA and a conventional downhill search is that SA also
spends some time exploring solutions with higher cost function, and in this way can climb
out of shallow valleys in the fitness function to explore other deeper
regions.

The total cost $ C_T $ will play the role of energy in the simulated annealing algorithm, so that the probability of accepting a change to a tree of cost $ C_T^i $, resulting in a tree of cost $ C_T^f $ is given as
\begin{equation}
P_i^f = \left\{ \begin{array}{rl}
 \exp{\left(-\frac{\Delta C_T}{T}\right)} &\mbox{ if $\Delta C_T > 0$} \\
  1 &\mbox{ otherwise}
       \end{array} \right.
\end{equation}
Where $P_i^f$ is the probability of going from state $ i $ to state $
f $, $ \Delta C_T = C_T^f - C_T^i $ is the change in the cost function
associated with going from state $ i $ to $ f $, and $ T $ is the
simulated annealing temperature parameter (not be confused
with ambient temperature). The small probability to accept a
higher cost tree during update allows the tree to climb out of
local valleys in the cost function. The algorithm proceeds by
making changes to the tree structure, calculating the change in cost
function, and then either accepting or rejecting the change by
comparing $P_{i}^{f}$ to a random number between 0 and
1. $T$ starts large and is reduced slowly. If $T$ has been reduced
sufficiently slowly, then the global minimum of the cost function is guaranteed to
be reached. In practice, the problem
space is too large to achieve this in reasonable time, and slightly
different trees with very similar cost are found if the algorithm is run with several random number seeds. The
most important consideration is the lowest achieved cost. As such, if the structure of trees generated varies between different
runs, we always display data from the run with the lowest cost function. As computational power increases,
longer runs will be achievable leading to progressively better
optimisations. The highest $T$ used here is $1\times 10^{10}$, dropping
during the algorithm to $10^{-5}$.  Typically a tree containing 1000
nodes will need $10^9$ updates, with a doubling of nodes taking
roughly quadruple the number of updates (up to around 6000 Nodes with 1 month of CPU time). The
large value of $A_{s}$ means that the supply of tissue is determined
by downhill search, while all other costs are minimised by simulated
annealing.

\subsection{Exploring the tree structure: Translations and node swaps}

The SA algorithm must have access to set of updates which allow it to alter the configuration of the tree. It is necessary to find changes that
can be made to the topological and geometrical structure of the tree
such that all possible solutions, between perfectly symmetric
structures and a single trunk vessel can be explored (i.e. the algorithm is ergodic). This is achieved by
allowing: (1) repositioning of bifurcations, which is achieved by translating a node in space (Fig. \ref{fig:samoves}a) and (2) swapping the
parent vessels of bifurcations between different parts of the tree (Fig. \ref{fig:samoves}b).
For all nodes but the root node, this move is valid, and performed consecutively it allows all possible tree topologies to be explored. If one of the two nodes is a direct parent of the other (i.e while traversing up the tree from one of the chosen nodes, the other node is encountered) the move is rejected to avoid forming a closed loop. With these two updates, the entire
parameter space of the tree can be explored, allowing the algorithm the opportunity to
reach a globally optimal solution.

\begin{figure}
\includegraphics[width=0.45\textwidth]{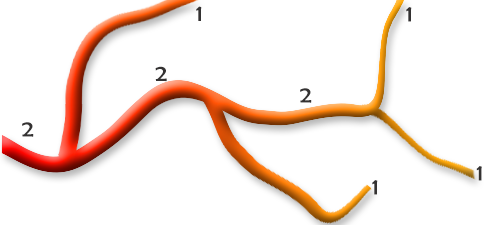}
\caption{Schematic of the Strahler ordering process.}
\label{fig:strahlerorder}
\end{figure}

\subsection{Strahler Order}

The Strahler (or stream) ordering method was first introduced to classify river systems, but can be applied to any bifurcating system. In standard Strahler ordering, nodes at the end of a tree (in this case the arterioles) are assigned a number 1. At a bifurcation, if two vessels (segments) of the same order meet, then the order of the parent vessel is 1 higher. However, if two vessels of different orders meet, the artery supplying these vessels has the largest order of the two. For example, if two arteries of order 1 meet, then the vessel supplying these arteries has order 2. If an artery of order 3 meets an artery of order 2, then the vessel supplying these arteries has order 3 (an example is shown in Fig. \ref{fig:strahlerorder}). Therefore, within this scheme, vessels with the lowest order are arterioles. The major vessels have the largest order. The Strahler order used here is then diameter adjusted following the approach in Ref. \cite{Jiang1994}.

Within the Strahler
ordering scheme it is possible to identify continuous sections of
vessels with the same order number. These are refereed to as
elements, so a single arterial element may pass through multiple
bifurcations. Throughout this article it is the properties of elements
which will be calculated for direct comparison with
Ref. \cite{Kassab2006}. We note that due
to the early termination of the simulated trees, calculated order
numbers are modified so that the root nodes have an order number equivalent
to that of the largest arteries of real coronary arterial trees. For example,
in the work of Kassab, the largest diameter defined Strahler order
number is 11, corresponding to the input artery. For a computer generated tree of only 6000 nodes spanning order
numbers 1-6, 5 must be added to each order number so that the orders of the root
nodes (largest vessels) match and a direct comparison can be made. This is consistent with assuming that the smallest vessels in the computer generated tree correspond to vessels of order 6. Which is due to the absence of smaller vessels downstream of the smallest arteries in the in-silico model.

\begin{figure*}[t]

  \centering
    \includegraphics[width=0.8\textwidth]{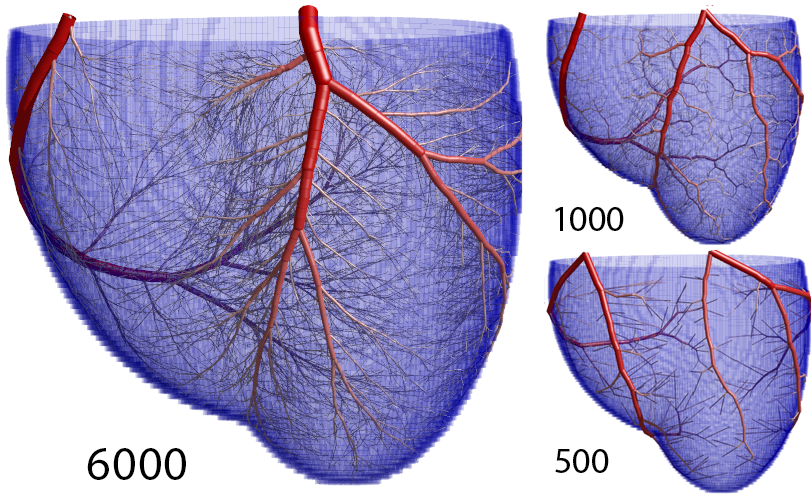}

    \caption{Images showing arterial trees grown with the
      approach detailed here. The number of terminal arterioles is increased from 500 to
      6000 (the total number of arterial segments is roughly twice this). There is consistency in the positioning of the
      larger arteries between the numerical method and the typical
      arrangement of the major arteries, suggesting that the coronary
      arteries may be the result of a biological process seeking the
      global minimum in metabolic demand.}
    \label{fig:multree}

\end{figure*}

\begin{figure*}[t]

  \centering
  \subfigure[][]{
    \includegraphics[width=0.35\textwidth]{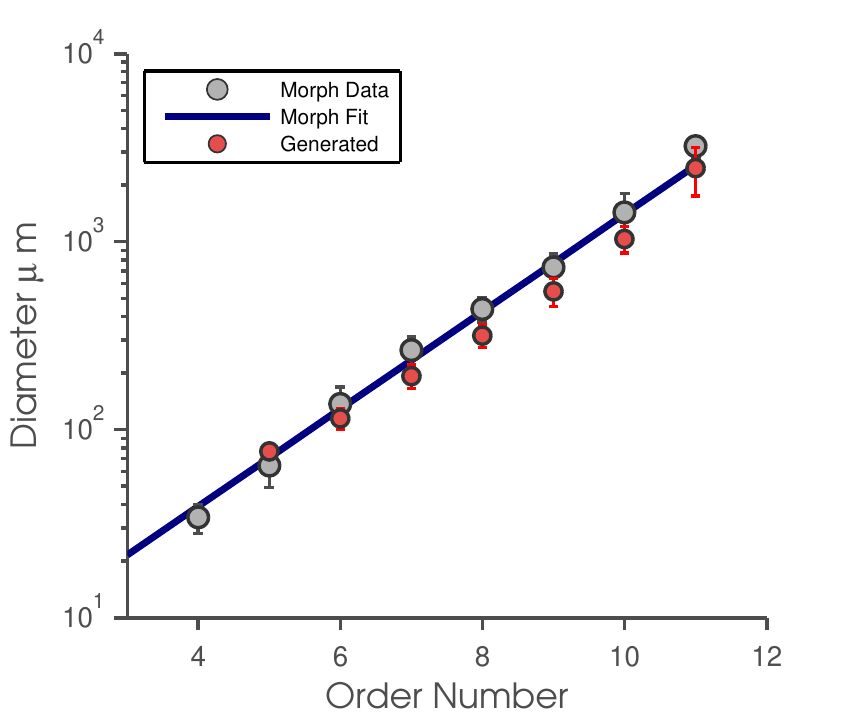}
    \label{fig:lengthsanddiametersa}
    }
    \subfigure[][]{
    \includegraphics[width=0.35\textwidth]{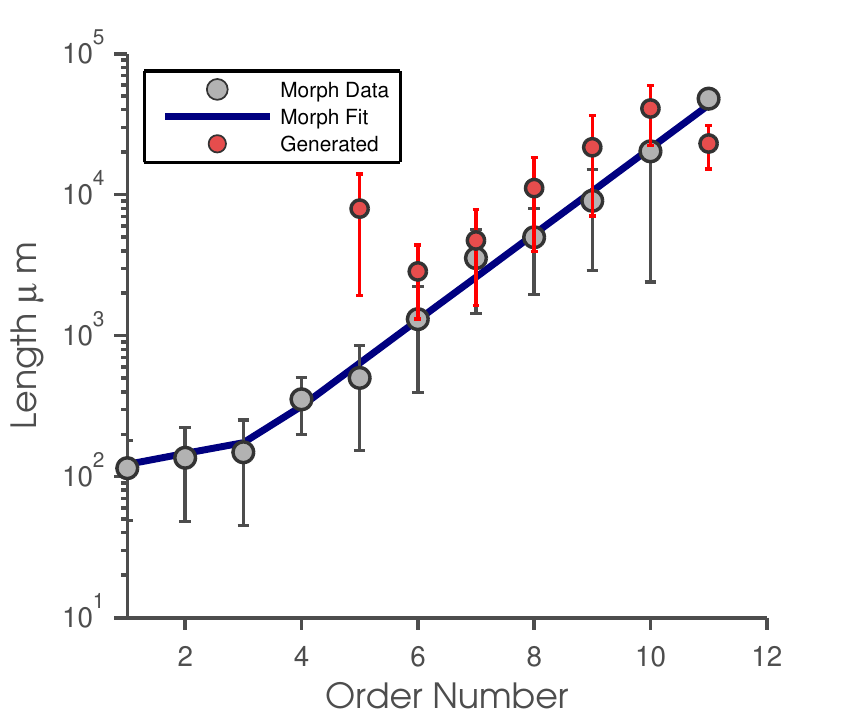}
    \label{fig:lengthsanddiametersb}
    }\\
    \subfigure[][]{
      \includegraphics[width=0.35\textwidth]{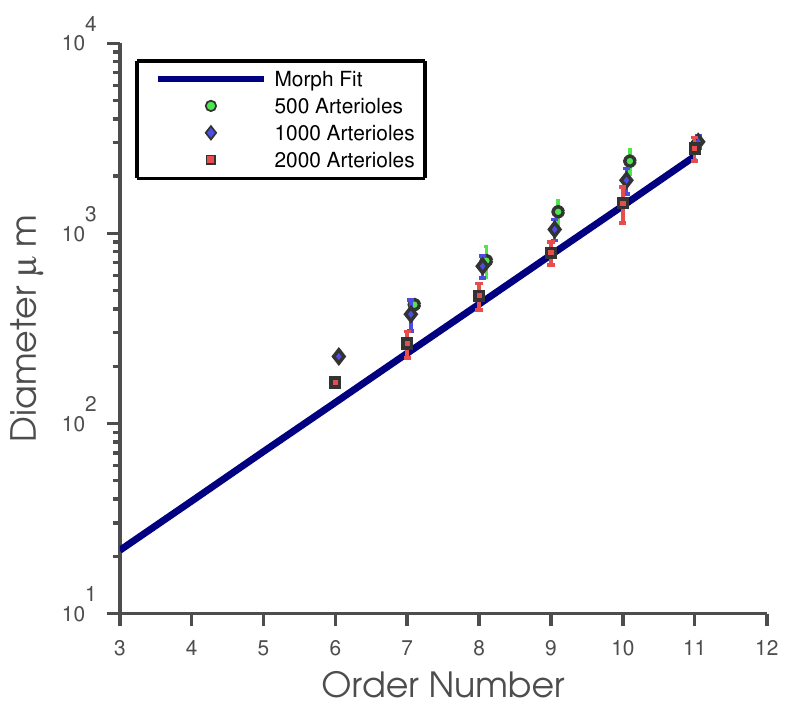}
      \label{fig:comprad}
    }
    \subfigure[][]{
      \includegraphics[width=0.35\textwidth]{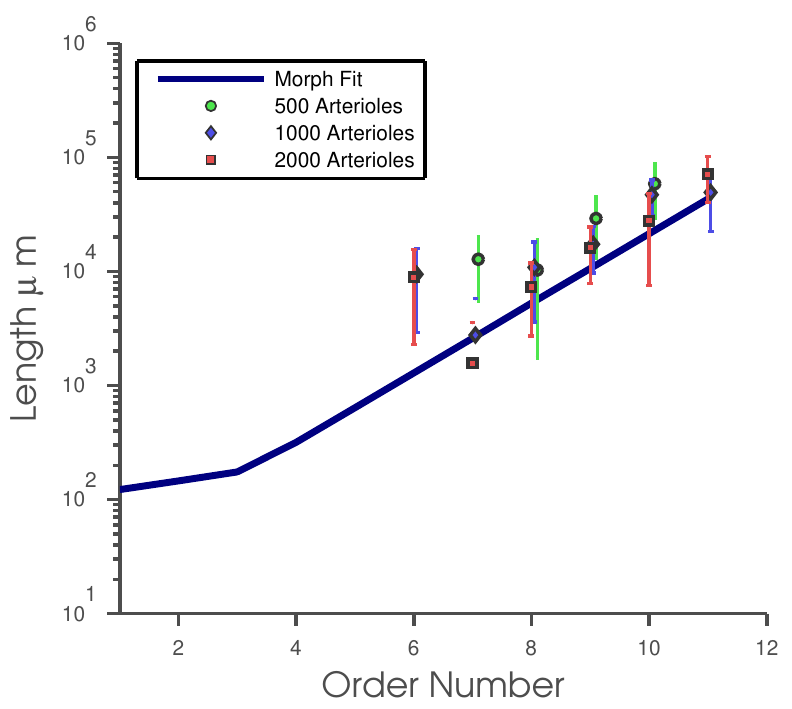}
      \label{fig:complen}
    }\\
    \caption{(a) Vessel diameter as a function of order in a tree with 6000 arterioles. Excellent agreement is found for vessels on all length scales. (b) Vessel length as a function of order number. Agreement is excellent for the major vessels (large order). The large variation seen for arterioles (lower order) is a result of early termination. Also shown are the morphological data reproduced from Table 2 of Ref. \cite{Kassab1993} for easy comparison. (c) and (d) are as (a) and (b), but for smaller trees to highlight the trend towards the morphological data as tree size increases. (Error bars show standard errors, both axes are logarithmic.)}
\end{figure*}

\begin{figure*}

  \centering
        \subfigure[][]{
    \includegraphics[width = 0.35\textwidth]{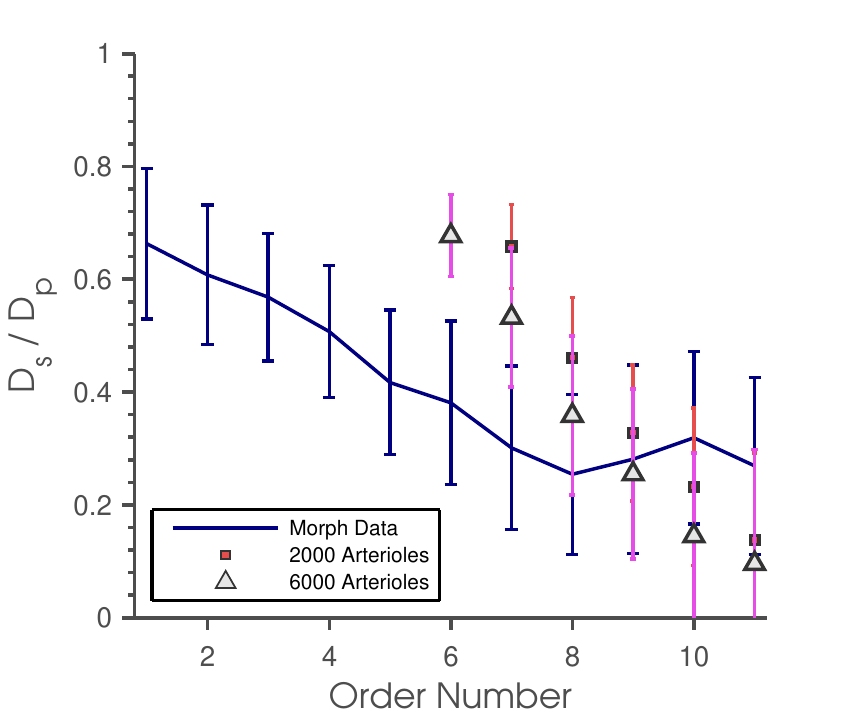}
    }
          \subfigure[][]{
    \includegraphics[width=0.35\textwidth]{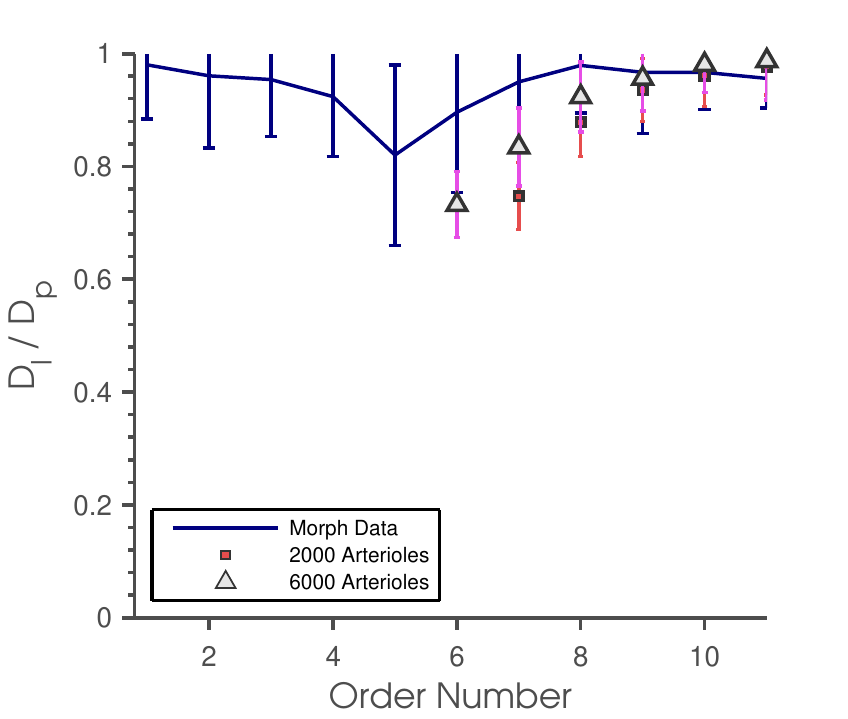}
    \label{dldmnumnodes}
    }\\

    \caption{The ratio of daughter vessel diameters ($D_{s}$ and
      $D_{l}$ are the diameters of the smallest and largest daughter
      vessels respectively) to diameters of parent segments, $D_{p}$
      as a function of order number, showing how the tree tends
      towards more symmetric branching at lower orders. Agreement with
      morphological data reproduced from tables in the online supplement of Ref. \cite{Kaimovitz2008} is good, if the early
      termination of the generated trees is taken into account, with
      the trend towards the morphological data as the tree size
      increases. Both graphs demonstrate that there are large trunks
      at high orders with the largest daughter vessel (panel (b)) of similar size to the
      parent vessel and another side artery which is much smaller (panel (a)). At
      smaller orders, the ratio becomes similar showing that the
      branchings of the smaller arteries are near symmetric. Realistic branching asymmetries are a clear advantage over other methods of generating arterial trees
      \textit{in-silico}.
       }
    \label{fig:multgraph}
\end{figure*}

\begin{figure*}[t]
     \includegraphics[width=0.8\textwidth]{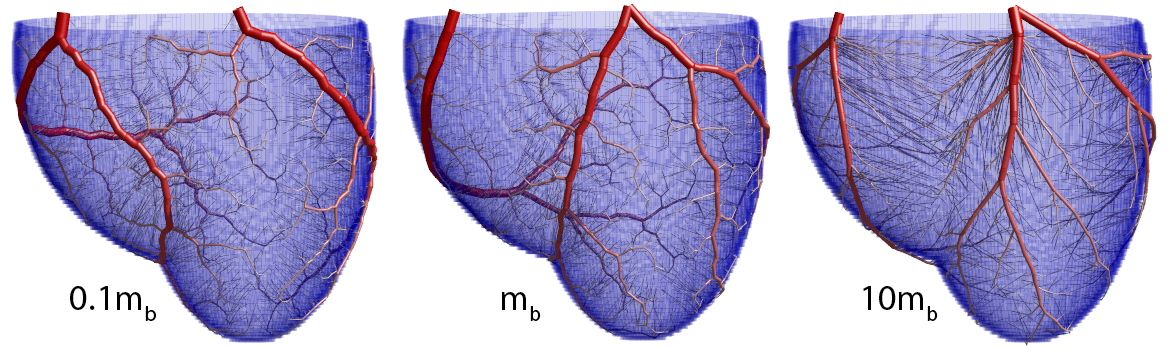}
\caption{Example trees generated with different values of $m_b$, which changes the relative weight of the pumping power to cost of maintaining blood in the optimisation. For small $m_{b}$ (corresponding to small hearts), vessels in the trees wind around - this is because there is little penalty to make a single wide vessel that curves to supply blood, rather than bifurcating. For large $m_{b}$ (corresponding to large hearts) the vessels travel as straight as possible.}
    \label{fig:metimg}
\end{figure*}

\begin{figure*}[t]

  \centering
  \subfigure[][]{
   \includegraphics[width=0.35\textwidth]{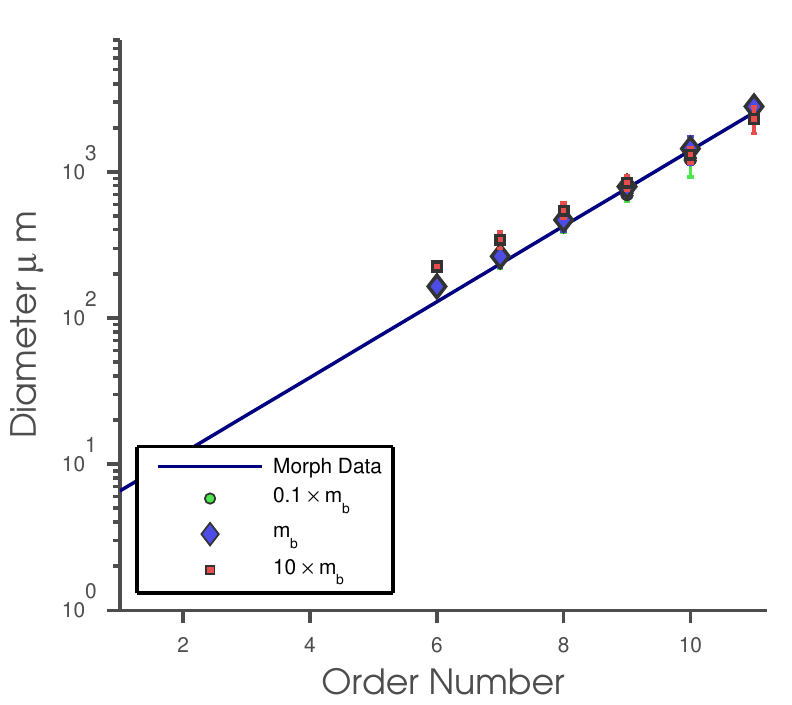}
    \label{metratrad}
    }
    \subfigure[][]{
   \includegraphics[width=0.35\textwidth]{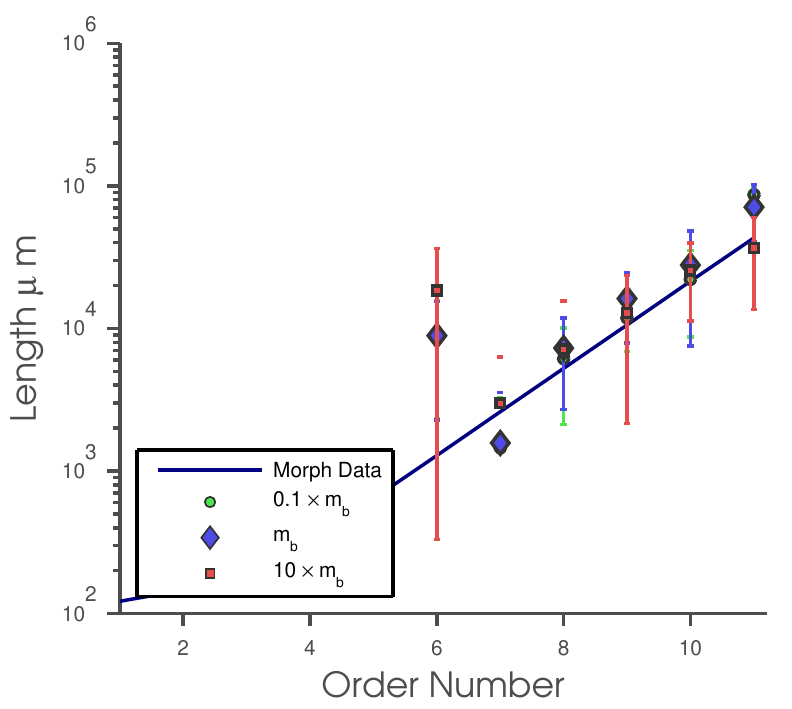}
    \label{metratlen}
    }
    \subfigure[][]{
   \includegraphics[width=0.35\textwidth]{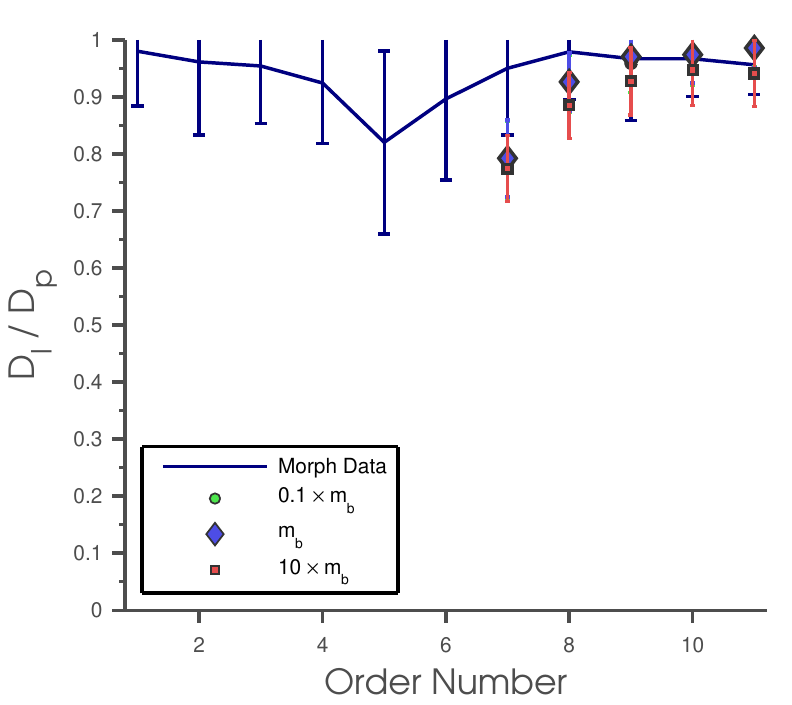}
    \label{dldmpv}
    }
    \subfigure[][]{
   \includegraphics[width=0.35\textwidth]{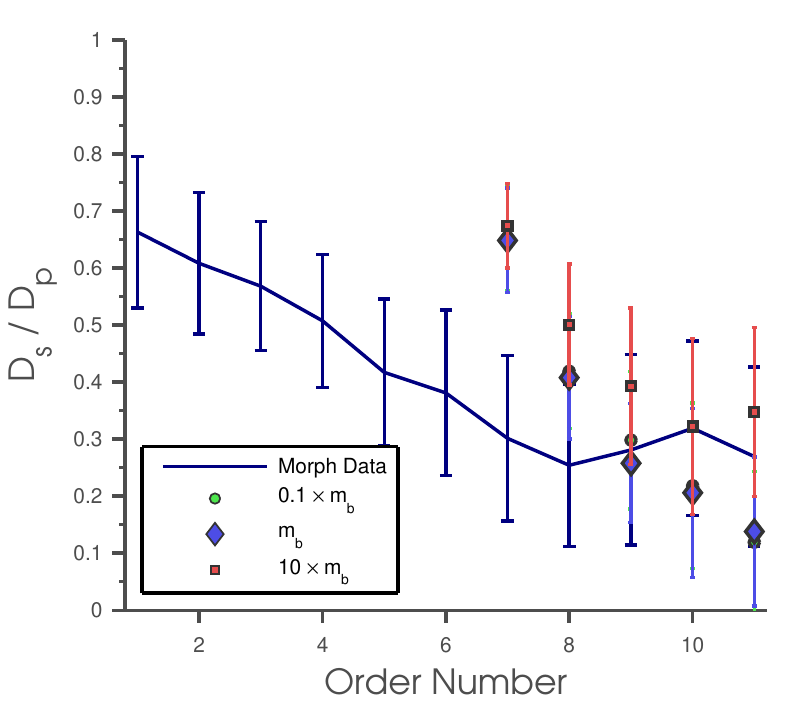}
    }
    \caption{(a) Diameter as a function of Order Number for trees with
      1000 vessels. Decreasing $m_b$, which
      describes the relative energy cost of an amount of blood and the
      power required to pump it, has little effect on the agreement of
      the diameters with morphological data. (b) For lengths however
      there is an obvious effect in the larger arteries, with regimes
      of high pumping cost being more accurate. The primary
      optimisation for high pumping cost then is to increase the
      length of the largest arteries. (c) and (d) The main effect is a change in the asymmetry of the branching of the largest arteries - for large $m_b$, the branches are more symmetric than for small $m_b$. As $m_b$ becomes very small, the limiting behaviour is broad trunks that wind around all the tissue, with a large number of very small offshoots that supply blood in the direct vicinity of the large vessel.}
\label{fig:metratio}
\end{figure*}

\section{Results}
\label{sec:results}

In this paper globally optimised vessels are grown using an SA based approach to supply a myocardial substrate, and validated through comparison with morphological data from the porcine arterial tree. We choose to examine the heart vasculature, since the structure of the large coronary arteries has been found to be similar between individuals
\cite{Glenny2007} and the full arterial tree has been well characterised in porcine models \cite{Kassab2006}. For modelling the coronary arteries we used the following parameters:
(1) A tissue substrate representing an ellipsoidal human heart muscle of mass 218g, constructed based on physiological parameters \cite{VanDenBroek1980} . The right ventricle was assumed to take the form of a super ellipsoid of exponent 2.5 and the left ventricle was represented by a simple ellipsoid.
Truncation of the ellipsoidal substrate was chosen so that the mass of the tissue corresponded to a reasonable physiological value given morphological data for ventricle thickness.
(2) Blood flow through each of the terminal segments of the tree was assumed to be constant, with each arteriole supplying an equal volume of tissue and homogeneous perfusion throughout the tissue parenchyma \cite{Pries2009}. These assumptions greatly simplify fluid dynamical calculations for estimating the total power needed to pump blood through the tree.
(3) The metabolic cost of maintaining a given volume of blood was assumed to be $ 641.3 J \mbox{ s}^{-1} $ per metre cubed of blood \cite{Liu2007} . For convenience, each arteriole supplies a sphere of tissue with a size calculated by assuming a mean blood flow per unit mass for cardiac muscle of $ 0.8 \mbox{ ml } \mbox{min}^{-1} \mbox{
  g}^{-1}$\cite{Frca2005}. The value taken from the literature was chosen such that it lay within the given error, but also conformed reasonably with both the ellipsoidal heart model, input flow and radii.
(4) The larger arteries with diameters greater than $ 0.01 \mbox{mm} $ were constrained to avoid penetration of the outer layer of heart tissue. This simplification differs slightly from real coronary vasculature, where progressive intrusion of arteries into the myocardium can be observed \cite{Sunni1986}. However, as the major arteries modelled by our method are far larger than the intra-myocardial vessels, a sharp cut-off is thought to provide a reasonable approximation.
(5) The starting positions of the two root arteries were fixed with a total input flow of $4.16^{-6} \mbox{
  m}^3 \mbox{ s}^{-1}$\cite{Johnson2008}. Relative radii of the two inputs to the tree were constrained via $r_{1}^{2.1}+r_{2}^{2.1}=[2.1 \mathrm{mm}]^{2.1}$ however, the relative sizes of root arteries and division of perfusion territories are determined by the method alone.
(6) The branching exponent varies throughout the coronary arterial tree, but for the larger arteries its value remains in the range 1.8 to 2.3. A variable branching exponent would greatly increase the computational cost of the approach, so a compromise value of 2.1 was chosen for the entire tree \cite{Kaimovitz2008}.

Coronary arterial trees containing increasing total numbers of vessels
grown using the SA based method are presented in
Fig. \ref{fig:multree}. In real human
coronary trees, there are 3 identifiable main coronary arteries (see
e.g. the schematic from Ref. \cite{openstax2013}): Left Anterior
Descending (LAD), Right Cardiac Artery (RCA) and Left Circumflex Artery
(LCX). The positions and relative dimensions of these are similar in
most humans, with major variations observed in less than 1\% of
healthy individuals \cite{Gharib2008}.  Trees grown using SA
(Fig. \ref{fig:multree}) adhere well to this structure. There is a
consistency in the placement of the larger arteries, although the RCA
appears slightly lower, and the right marginal artery appears slightly
shorter, in our models. Overall, visual inspection of the arterial
structure appears extremely promising.

To provide a quantitative comparison of our trees with anatomical data, the topological characteristics of the computer generated coronary artery trees were extracted and compared to morphological data characterising the pig coronary arteries published by Kassab {\it et al.} \cite{Kassab2006} Kassab and colleagues used a combination of corrosion casting and optical sectioning to obtain detailed morphometric data, tabulated using the Strahler (or stream) ordering scheme to denote elements of the tree of varying scale. Within this scheme, the lowest Strahler order numbers correspond to the smallest arterioles and the largest numbers refer to major vessels (for details on Strahler ordering see method). To directly compare arterial diameters, lengths, and branching properties, of our computer-generated arterial tree with real data from pig coronary arteries, averages were obtained over all elements of the same diameter defined Strahler order.

The mean vessel diameters are shown as a function of order number, for a tree comprising of 6000 arterioles (12000 vessel segments) in Fig. \ref{fig:lengthsanddiametersa}. Excellent agreement is found between the trees generated {\it in-silico} and the morphological data. Only slight deviations from the morphological data can be seen for the smallest vessels (lowest order arteries) in the generated tree. This is likely to be due to the combination of integer order numbers and the condition that terminal sites are of constant radius. The result of this constraint is that the terminal radii will only match the anatomical data for a correct choice of the number of arterioles. Fig \ref{fig:comprad} shows the effects on diameter of increasing the number of arterioles from 500 to 2000. Agreement is generally good, regardless of the number of terminal arteries, and there is a clear trend towards matching the experimental data as simulated tree size increases.
	Fig \ref{fig:lengthsanddiametersb} compares average vessel length in the model and porcine morphological data as a function of order number. For the largest arteries (high order numbers) the agreement is excellent. Although the lengths of the smaller arteries (Strahler orders  $<$ 7) in the computer generated tree tended to be overestimated, this can be easily explained by the fact that the smallest vessels are required to bridge a gap that would normally be filled by inclusion of lower order vessels in a larger simulation. As the number of generated vessels is increased, the agreement with morphological data improves (Fig. \ref{fig:complen}).

	Previously, the best methods available for the computer generation of arterial trees struggled to recreate realistic branching asymmetry. Fig \ref{fig:multgraph} shows the ratio of daughter to mother vessel radii for the largest and smallest daughter vessels as a function of order number. This provides a measure of the branching asymmetry of the tree, where small ratios indicate that branching is symmetric, while ratios approaching 1 suggest a large trunk vessel with small branches. For Strahler orders corresponding to microvascular arterioles, both the computer generated and true morphology approach 0.7, which is consistent with perfectly symmetric branching where both daughter vessels are of similar size. Agreement with the morphological data from Ref. \cite{Kaimovitz2008}  improves as the size of the computer generated tree increases. This is not the result of any special input parameters or initial conditions. The trees are topologically and spatially randomised before SA optimisation begins, and are allowed to explore the entire parameter space during optimisation. The observed asymmetry is purely the result of a balance between pumping power and metabolic maintenance cost, and is a major improvement in predicting the trunk-like structure of major vessels.

Our final figures show the effect of altering the metabolic energy cost of blood per unit volume $m_b$. The largest morphological change is found in the lengths of the larger arteries (Fig \ref{fig:metratio}). As $m_b$ increases, bifurcation symmetry is also increased in the larger arteries and as a result there is an increase in the number of Strahler orders present in the tree (Fig \ref{fig:metratio}). The explanation for these scaling behaviours is evident when considering the limiting cases. For $m_b = 0$ the power involved in pumping the blood dominates the optimisation, which leads to a large, `snaking' artery with small side branches that supply the tissue. This large artery would cover the entire surface of the heart, and the configuration is equivalent to a completely asymmetric binary tree. For a large $m_b$ value (or small power cost) there is a huge penalty associated with larger arteries, and so their lengths are contracted. In order to accommodate the reduction in length, the larger arteries must bifurcate more frequently and symmetrically. Additionally the high volume cost causes the trunk artery to minimise its total length, resulting in a much straighter path across the tissue. Less extreme examples of this behaviour can been seen in Fig \ref{fig:metimg}, with meandering arteries for small $m_b$ and straight arteries for large $m_b$.

The change in $m_b$ can also be interpreted as a change in length scale as follows: Once the large vessels have been excluded from the tissue and all tissue is supplied, the remaining cost function that is optimised has the form,
\begin{equation*}
C=m_{b}\pi r^2 l + \frac{8\mu l Q^2}{\pi r^4}
\end{equation*}
now, make the transformations, $r\rightarrow r'=Ar$, $l\rightarrow l'=Al$. Then the cost function becomes,
\begin{equation*}
C=A^3m_{b}\pi r^2 l + \frac{8\mu A l Q^2}{A^4 \pi r^4}
\end{equation*}
since the optimum in the cost function is the same independent of a multiplicative factor that acts on all terms, then we can absorb a factor of $1/A^3$ into the cost function to obtain:
\begin{equation*}
C'=A^6m_{b}\pi r^2 l + \frac{8\mu l Q^2}{\pi r^4}
\end{equation*}
identifying a new $m_b'=A^6 m_b$ the cost function now has the same form. Since changing $m_b$ is equivalent to changing the length scale, these results suggest that there are likely to be structural
differences between species of different sizes, as the power required
to pump blood becomes relatively more important than the metabolic
demand to maintain blood volume in small vessels. In the absence of morphological data, visual comparison of the coronary arteries tentatively indicates that vessels meander around in smaller species \cite{Yoldas2010} and that vessels are straighter in larger species \cite{ozgel2004}.

\section{Discussion}
\label{sec:conclusions}

We have developed a powerful and universal method for growing
arterial trees \textit{in-silico}, which is capable of identifying the
near globally optimal configuration of arteries for arbitrarily shaped
tissues with heterogenous blood supply demands. As input, the method only needs information about the tissue structure and the entry point positions of the largest arteries. From this information, the approach generates morphologically and structurally
accurate coronary arterial trees at almost every length scale. This is
a significant improvement on previous optimisation methods, which
failed to reproduce the consistent structure found in the coronary
arteries. We have shown that the method improves with the number of
vessels modeled, so that, as computing power increases, there is a systematic improvement in the
accuracy of the generated trees. To our knowledge, no other method can
generate realistic arterial trees that closely match morphological
data by taking only the shape of the tissue as input, and claim
systematic improvement in the generated trees with increased
computational power.

We expect that our method could have several useful applications. Our first application is to use cardiac and partial cerebral vasculature structures (e.g. MCA territory) computed using this method as input to models of embolic stroke and other infarctions, since these models require detailed vasculature structure over a range of length scales which are not available to imaging techniques  \cite{chung2007,hague2009,hague2013}. This will require further validation of the algorithm against cerebral arterial data. Such vasculature would be downstream of the Circle of Willis, a source of major anatomical variation and likely not a structure reproducible by the current algorithm. Computational models of stroke combined with doppler ultrasound have potential to provide further information regarding embolic burden during major operations \cite{chung2015}.

There are several other potential applications. Models of arterial trees generated by our method may help to
improve the interpretation of medical images though advanced image segmentation techniques. Vessels identified  through automatic segmentation techniques can be connected via algorithms such as the one presented\cite{Jiang2010}. In addition, segmentation can be limited to the location of a reduced set of bifurcation points, and the algorithm used to fill in any missing vessels which connected them \cite{Suetens2009}.

We also speculate that the algorithm could be of use in designing the structures of vasculature for artificial tissues. Once the very difficult and intricate process of making networks of vessels in artifical tissue has been achieved\cite{Novosel2011, Kim2013, Baranski2013}, the opportunity to optimise or inform their design will be available. A common problem during the growth of artificial tissue is that regions of cells can die due to lack of nutrients and oxygen. For instance an artificial skin graft may be optimised for increased healing, by having its vasculature designed such that there is higher overall perfusion with minimal loss of useful tissue. Even more speculatively, artificially grown organs may have a vasculature designed to minimise their impact on the cardiovascular system.


\section{Acknowledgments}	

We thank Chloe Long, Martin Bootmann and Uwe Grimm for useful discussions.

\section{Data Accessibility}
Data presented in this articles can be found in a compressed zip folder as Electronic Supplementary material. The folders contain MATLAB scripts capable of plotting the data.

\section{Competing Interests}
We have no competing interests.

\section{Authors' Contributions}
JK developed the algorithm, acquired, analysed and interpreted data, and contributed to drafting the article. EC co-conceived the study, co-supervised the project and contributed to drafting the article. JPH conceived the study, developed initial versions of the algorithm, contributed to the analysis and interpretation of data, supervised the project, and contributed to drafting of the article. All authors gave final approval for publication.

\section{Funding}
JK acknowledges EPSRC grant EP/P505046/1. EMLC acknowledges support from a British Heart Foundation Intermediate Basic Science Research Fellowship (FS/10/46/28350).

\appendix
\section{Convergence and Consistency}
The primary purpose of the algorithm is to produce arterial tree configurations which conform to those found in living organisms. As the algorithm itself relies only upon optimisation principles, the close agreement with experimental results implies an evolutionary pressure towards a structure with minimal power consumption. This is itself a far from new concept, however in this paper we have shown that energetic constraints lead not only to a morphometrically realistic tree, but also to the production of major arteries which closely follow the paths of major arteries in living systems.

Whilst it is clear that the algorithm produces both morphometrically and geometrically realistic structures, what is not clear is how close the optimisation procedure gets to the global energy minimum, or indeed whether there is a non-degenerate energy minimum at all. For any given topological configuration there is a single, non-degenerate energy minimum which it is possible to approach using Newton-Rhapson (provided the solution space is convex, which would not be true for more complex structures). In contrast, the topological space for any even modestly sized tree is huge, highly degenerate and not easily searchable. Even if one excludes degenerate topological structures, which in the case of the swap node procedure outlined earlier would imply never swapping two nodes with the same number of distal terminal sites, the number of possible configurations is still massive. It is entirely possible that two distinct topological configurations share a degenerate energy level, and proving that this is not the case appears difficult.

While it may be that the global energy minimum is highly degenerate, the algorithm itself can still be characterised in terms of reliability and convergence. For reliability, we can perform visual inspections on trees and assess their similarity. It must be noted here that the geometry in which the tree is grown will have a large effect on the consistency of the results. For instance, in the case of a circular section of tissue with an input in the centre, there is a high degree of rotational symmetry. In the case of convergence, we can produce many trees and plot the frequency distribution of their resultant energies, or as in this case the average and variance of the cost as a function of SA steps.

\begin{figure*}[t]
    \includegraphics[width=0.5\textwidth]{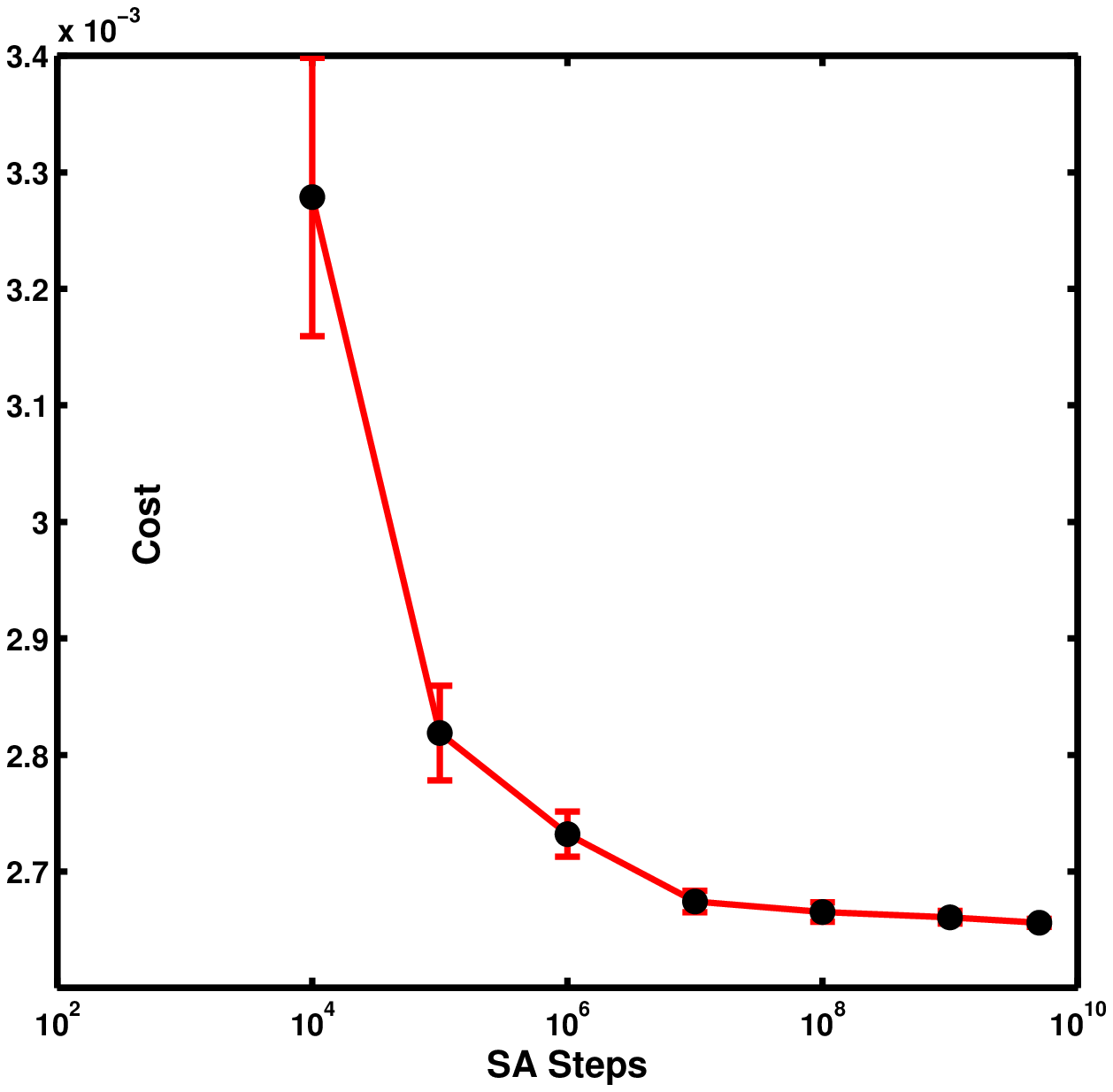}
    \caption{Average cost and standard error vs SA steps for a 127 node tree grown in a 2D plane. }
    \label{fig:energyvssasteps}
\end{figure*}

The convergence and consistency test trees were generated on a 2D plane with the input placed in one corner. The trees consisted of 127 nodes total (64 end nodes) and had a bifurcation exponent of $3.0$. The 2D tissue plane was sized at 10cm by 10cm and the root radius at 2.4mm. Each tree was optimised for a given number of simulated annealing steps, with the minimum energy of the SA run being recorded. The average energy reached for a given number of SA steps was then calculated (Fig. \ref{fig:energyvssasteps}). The results show a clear trend towards lower average energy and standard deviation as the number of SA steps increases. The high variance at the lower numbers of SA steps are typical of a system which has been quenched, i.e high temperature disorder has been locked into the system, which had not had sufficient time to reach equilibrium.

\begin{figure*}[t]
  \centering
  \begin{tabular}{c | c | c | c}
    $10^4$ Steps & $10^6$ Steps & $10^8$ Steps & $5 \times 10^9$ \\ \hline

    \includegraphics[width=0.2\textwidth]{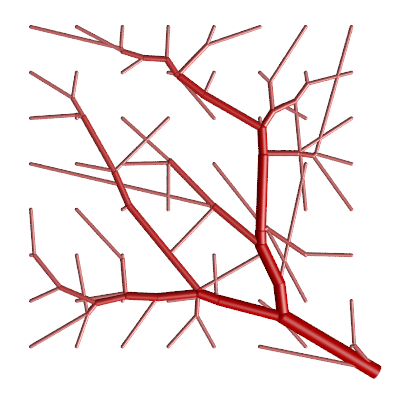} &
    \includegraphics[width=0.2\textwidth]{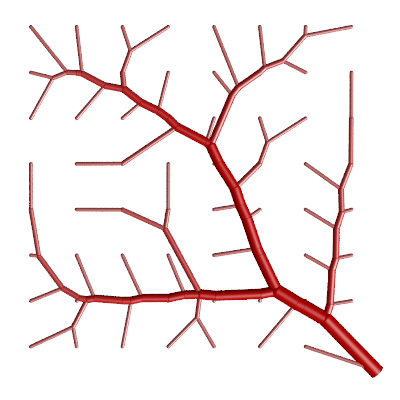} &
    \includegraphics[width=0.2\textwidth]{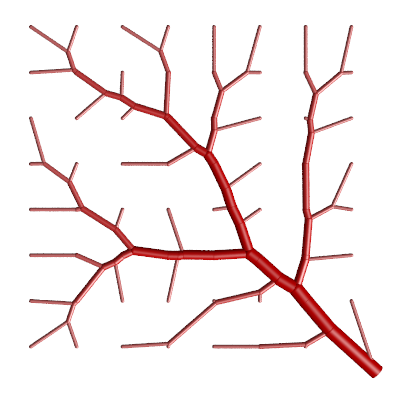} &
    \includegraphics[width=0.2\textwidth]{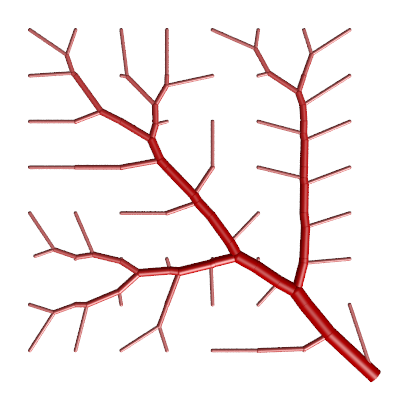} \\

    \includegraphics[width=0.2\textwidth]{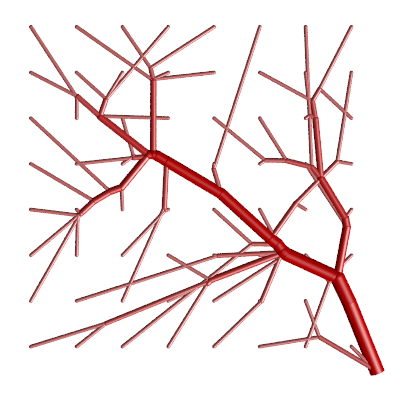} &
    \includegraphics[width=0.2\textwidth]{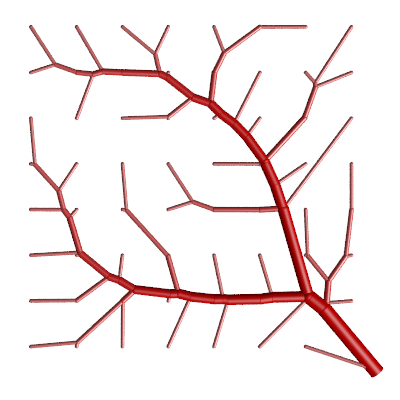} &
    \includegraphics[width=0.2\textwidth]{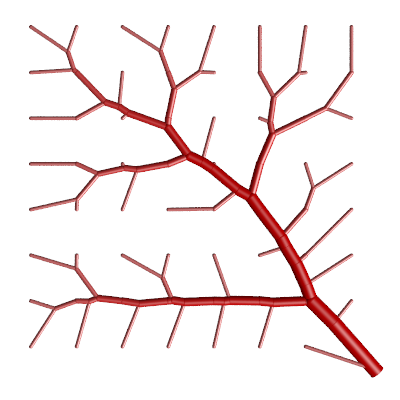} &
    \includegraphics[width=0.2\textwidth]{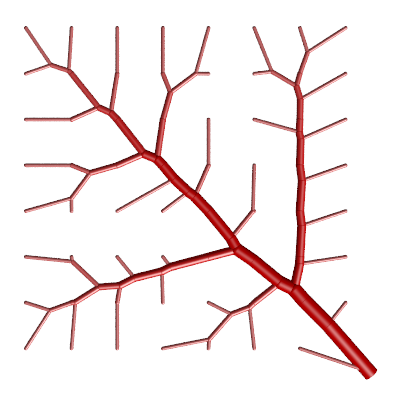} \\

    \includegraphics[width=0.2\textwidth]{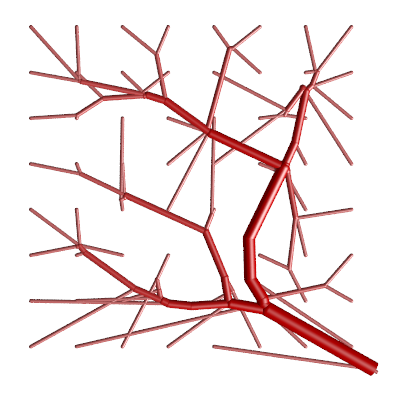} &
    \includegraphics[width=0.2\textwidth]{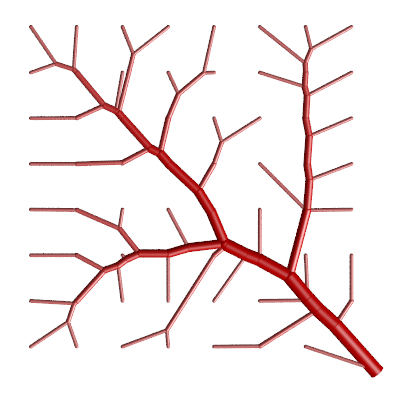} &
    \includegraphics[width=0.2\textwidth]{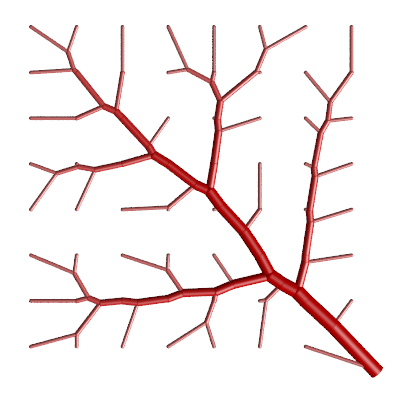} &
    \includegraphics[width=0.2\textwidth]{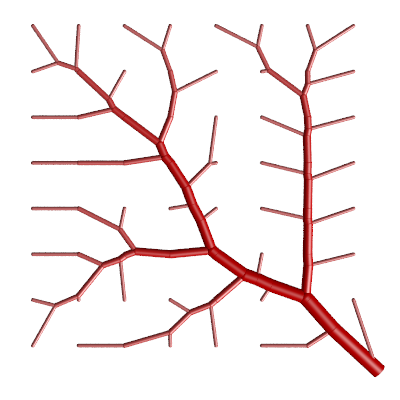} \\

  \end{tabular}
  \caption{Example of trees grown for various different numbers of SA step numbers.}
  \label{fig:consistency}
\end{figure*}

For the consistency test we have a produced Fig. \ref{fig:consistency}, which show trees generated for three different numbers of steps. As would be expected, at low numbers of SA steps the trees are very dissimilar, however as the number of steps is increases the similarity between the overall trees increases dramatically, with a main diagonal artery dominating the structure.

\bibliographystyle{unsrt}

\bibliography{references}

\begin{thebibliography}{10}

\bibitem{ZamirBook}
M.~Zamir.
\newblock {\em The Physics of Pulsatile flow}.
\newblock Springer, 2000.

\bibitem{Frame}
MDS Frame and IH~Sarelius.
\newblock {\em Microvascular research}, 50:301--310, 1995.

\bibitem{Murray1926a}
C~D Murray.
\newblock {\em Proceedings of the National Academy of Sciences of the United
  States of America}, 12(3):207--14, March 1926.

\bibitem{Rossitti1993}
S~Rossitti and J~L\"{o}fgren.
\newblock {\em Stroke: A Journal of Cerebral Circulation}, 24(3):371--377,
  1993.

\bibitem{Kassab1993}
G~S Kassab, C~A Rider, N~J Tang, and Y~C Fung.
\newblock {\em American Journal of Physiology}, 265(1):H350--65, 1993.

\bibitem{Cassot2010}
F~Cassot, F~Lauwers, S~Lorthois, P~Puwanarajah, V~Cances-Lauwers, and
  H~Duvernoy.
\newblock {\em Brain research}, 1313:62--78, February 2010.

\bibitem{Huo2012}
Y~Huo and G~S Kassab.
\newblock {\em Biomedical Engineering}, 9:190--200, 2012.

\bibitem{Huo2009}
Y~Huo and G~S Kassab.
\newblock {\em Biophysical journal}, 96(2):347--53, January 2009.

\bibitem{Kassab2006}
G~S Kassab.
\newblock {\em American journal of physiology. Heart and circulatory
  physiology}, 290(2):H894--903, February 2006.

\bibitem{Bengtsson2003}
H~U Bengtsson and P~Ed\'{e}n.
\newblock {\em Journal of theoretical biology}, 221(3):437--43, April 2003.

\bibitem{West1997}
G~B West, J~H Brown, and B~J Enquist.
\newblock {\em Science}, 276(5309):122--126, 1997.

\bibitem{Karch1999}
R~Karch, F~Neumann, M~Neumann, and W~Schreiner.
\newblock {\em Computers in biology and medicine}, 29(1):19--38, January 1999.

\bibitem{Karch2000}
R~Karch, F~Neumann, M~Neumann, and W~Schreiner.
\newblock {\em Annals of Biomedical Engineering}, 28(5):495--511, May 2000.

\bibitem{Kaimovitz2005}
B~Kaimovitz, Y~Lanir, and G~S Kassab.
\newblock {\em Annals of biomedical engineering}, 33(11):1517--35, November
  2005.

\bibitem{Schreiner1993}
W~Schreiner and P~F Buxbaum.
\newblock {\em IEEE transactions on bio-medical engineering}, 40(5):482--91,
  May 1993.

\bibitem{Kaimovitz2010}
B~Kaimovitz, Y~Lanir, G~S Kassab, S~Nees, G~Juchem, N~Eberhorn, M~Thallmair,
  S~F\"{o}rch, M~Knott, A~Senftl, T~Fischlein, B~Reichart, D~R Weiss, and H~G~M
  {Van Beek}.
\newblock {\em Am J. Phys. Heart Circ. Phys.}, 299(July 2010):1064--1067, 2010.

\bibitem{Mayrovitz1983}
H~N Mayrovitz and J~Roy.
\newblock {Microvascular blood flow: evidence indicating a cubic dependence on
  arteriolar diameter.}
\newblock {\em The American journal of physiology}, 245(6):H1031--H1038, 1983.

\bibitem{Nakamura2014}
Horsfield K. and Woldenberg MJ.
\newblock Diameter and cross-sectional areas of branches in the human pulmonary
  arterial tree.
\newblock {\em Anat Rec.}, 223(3):245--251, 1989.

\bibitem{Karch2000a}
R~Karch, F~Neumann, M~Neumann, and W~Schreiner.
\newblock {\em Annals of biomedical engineering}, 28(5):495--511, May 2000.

\bibitem{Schreiner1997}
W~Schreiner, F~Neumann, M~Neumann, R~Karch, A~End, and S~M Roedler.
\newblock {\em The Journal of general physiology}, 109(2):129--140, 1997.

\bibitem{Schreiner2006}
W~Schreiner, R~Karch, M~Neumann, F~Neumann, P~Szawlowski, and S~Roedler.
\newblock {\em Medical Engineering \& Physics}, 28(5):416--429, 2006.

\bibitem{Georg2010}
HorstK. Hahn, Manfred Georg, and Heinz-Otto Peitgen.
\newblock In Gabriele~A. Losa, Danilo Merlini, Theo~F. Nonnenmacher, and
  Ewald~R. Weibel, editors, {\em Fractals in Biology and Medicine}, Mathematics
  and Biosciences in Interaction, pages 55--66. Birkh\"{a}user Basel, 2005.

\bibitem{Kassab1997}
G~S Kassab, E~Pallencaoe, A~Schatz, and Y~C Fung.
\newblock {\em American Journal of Physiology}, 273(6 Pt 2):H2832--H2842, 1997.

\bibitem{Fung2011}
George S~K Fung, W~Paul Segars, Grant~T Gullberg, and Benjamin M~W Tsui.
\newblock Development of a model of the coronary arterial tree for the 4d xcat
  phantom.
\newblock {\em Physics in Medicine and Biology}, 56(17):5651, 2011.

\bibitem{ref1}
Holger Perfahl, Helen~M. Byrne, Tingan Chen, Veronica Estrella, Tomás
  Alarcón, Alexei Lapin, Robert~A. Gatenby, Robert~J. Gillies, Mark~C. Lloyd,
  Philip~K. Maini, Matthias Reuss, and Markus~R. Owen.
\newblock Multiscale modelling of vascular tumour growth in 3d: The roles of
  domain size and boundary conditions.
\newblock {\em PLoS ONE}, 6(4):e14790, 04 2011.

\bibitem{ref2}
A.R.A. Anderson and M.A.J. Chaplain.
\newblock Continuous and discrete mathematical models of tumor-induced
  angiogenesis.
\newblock {\em Bulletin of Mathematical Biology}, 60(5):857--899, 1998.

\bibitem{ref3}
Steven~R. McDougall, Alexander~R.A. Anderson, and Mark~A.J. Chaplain.
\newblock Mathematical modelling of dynamic adaptive tumour-induced
  angiogenesis: Clinical implications and therapeutic targeting strategies.
\newblock {\em Journal of Theoretical Biology}, 241(3):564 -- 589, 2006.

\bibitem{ref4}
Anusuya Das, Douglas Lauffenburger, Harry Asada, and Roger~D. Kamm.
\newblock A hybrid continuum{\textendash}discrete modelling approach to predict
  and control angiogenesis: analysis of combinatorial growth factor and matrix
  effects on vessel-sprouting morphology.
\newblock {\em Philosophical Transactions of the Royal Society of London A:
  Mathematical, Physical and Engineering Sciences}, 368(1921):2937--2960, 2010.

\bibitem{kirkpatrick1983a}
S.~Kirkpatrick, C.~D. Gelatt, and M.P. Vecchi.
\newblock {\em Science}, 220:671--680, 1983.

\bibitem{Liu2007}
Y~Liu and G~S Kassab.
\newblock {\em American journal of physiology Heart and circulatory
  physiology}, 292(3):H1336--H1339, 2007.

\bibitem{Vinnakota2004}
Kalyan~C Vinnakota and James~B Bassingthwaighte.
\newblock {\em American journal of physiology. Heart and circulatory
  physiology}, 286(5):H1742--H1749, 2004.

\bibitem{Uren1994}
N~G Uren, J~A Melin, B~{De Bruyne}, W~Wijns, T~Baudhuin, and P~G Camici.
\newblock {\em The New England journal of medicine}, 330(25):1782--1788, 1994.

\bibitem{Hall2010}
John~E Hall.
\newblock {\em {Guyton and Hall Textbook of Medical Physiology}}.
\newblock Elsevier, 2010.

\bibitem{bfoscillations}
R.B King and J.B. Bassingthwaighte.
\newblock Temporal fluctuations in regional myocardial flows.
\newblock {\em Pflugers Arch.}, 413:336--342, 1989.

\bibitem{Amanatides1987}
J~Amanatides and A~Woo.
\newblock {\em Delta}, i(3):3--10, 1987.

\bibitem{Henderson2003}
D~Henderson, S~H Jacobson, and A~W Johnson.
\newblock In {\em Handbook of metaheuristics}, chapter~10, pages 287--319.
  Kluwer, 2003.

\bibitem{Kirkpatrick1983}
S~Kirkpatrick, C~D Gelatt, and M~P Vecchi.
\newblock {\em Science (New York, N.Y.)}, 220(4598):671--80, May 1983.

\bibitem{Jiang1994}
Z~L Jiang, G~S Kassab, and Y~C Fung.
\newblock {\em Journal of Applied Physiology}, 76(2):882--892, 1994.

\bibitem{Kaimovitz2008}
B~Kaimovitz, Y~Huo, Y~Lanir, and G~S Kassab.
\newblock {\em American journal of physiology Heart and circulatory
  physiology}, 294(2):H714--H723, 2008.

\bibitem{Glenny2007}
R~Glenny, S~Bernard, B~Neradilek, and N~Polissar.
\newblock {\em Proceedings of the National Academy of Sciences of the United
  States of America}, 104(16):6858--6863, 2007.

\bibitem{VanDenBroek1980}
J.~J. J.~M. {Van Den Broek} and M.~H. L.~M. {Van Den Broek}.
\newblock {\em J. Biomechanics}, 13:493--503, 1980.

\bibitem{Pries2009}
A~R Pries and T~W Secomb.
\newblock {\em Cardiovascular Research}, 81(2):328--335, 2009.

\bibitem{Frca2005}
F.J. Klocke, I.L. Bunnell, D.G Greene, S.M. Wittenberg, and J.P. Visco.
\newblock {\em Circulation}, 50(3):547--549, 1974.

\bibitem{Sunni1986}
S~Sunni, S~P Bishop, S~P Kent, and J~C Geer.
\newblock {\em Archives of pathology laboratory medicine}, 110(5):375--381,
  1986.

\bibitem{Johnson2008}
K~Johnson, P~Sharma, and J~Oshinski.
\newblock {\em Journal of Biomechanics}, 41(3):595--602, 2008.

\bibitem{openstax2013}
OpenStax College.
\newblock {\em Anatomy and Physiology [ISBN 978-1-938168-13-0, Available at
  Connexions Web site, http://cnx.org/content/col11496/1.6/]}.
\newblock June 2013.

\bibitem{Gharib2008}
A.~M. Gharib, V~B Ho, D~R Rosing, D~A Herzka, M~Stuber, A~E Arai, and R~I
  Pettigrew.
\newblock {\em Radiology}, 247(1):220--227, April 2008.

\bibitem{Yoldas2010}
A~Yoldas, E~Ozmen, and V~Ozdemir.
\newblock {\em Journal of the South African Veterinary Association},
  81(4):247--252, 2010.

\bibitem{ozgel2004}
O.~Ozgel, A.~Haligur, N.~Dursun, and E.~Karakurum.
\newblock {\em Anat. Histol. Embryol}, 33:278--283, 2004.

\bibitem{chung2007}
Emma M~L Chung, James~P Hague, and David~H Evans.
\newblock {\em Physics in medicine and biology}, 52(23):7153--66, December
  2007.

\bibitem{hague2009}
JP~Hague and EML Chung.
\newblock {\em Physical Review E}, 80(5):051912, 2009.

\bibitem{hague2013}
J.P. Hague, C.~Banahan, and E.M.L. Chung.
\newblock {\em Phys. Med. Bio.}, 58:4581, 2013.

\bibitem{chung2015}
E.M.L.~Chung {\it et al.}
\newblock {\em PLOS one}, 10:e0122166, 2015.

\bibitem{Jiang2010}
Y~Jiang, Z~Zhuang, A~J Sinusas, and X~Papademetris.
\newblock {\em Conference on Computer Vision and Pattern Recognition Workshops
  IEEE Computer Society Conference on Computer Vision and Pattern Recognition.
  Workshops}, pages 178--185, June 2010.

\bibitem{Suetens2009}
Pieter Bruyninckx, Dirk Loeckx, Dirk Vandermeulen, and Paul Suetens.
\newblock {Segmentation of liver portal veins by global optimization}.
\newblock {\em Imaging}, 7624:76241Z--76241Z--12, 2010.

\bibitem{Novosel2011}
E.~C. Novosel, C.~Kleinhans, and P.~J. Kluger.
\newblock {\em Advanced Drug Delivery Reviews}, 63:300--311, 2011.

\bibitem{Kim2013}
Sudong Kim, Hyunjae Lee, Minhwan Chung, and Noo~Li Jeon.
\newblock Engineering of functional{,} perfusable 3d microvascular networks on
  a chip.
\newblock {\em Lab Chip}, 13:1489--1500, 2013.

\bibitem{Baranski2013}
Jan~D. Baranski, Ritika~R. Chaturvedi, Kelly~R. Stevens, Jeroen Eyckmans, Brian
  Carvalho, Ricardo~D. Solorzano, Michael~T. Yang, Jordan~S. Miller,
  Sangeeta~N. Bhatia, and Christopher~S. Chen.
\newblock Geometric control of vascular networks to enhance engineered tissue
  integration and function.
\newblock {\em Proceedings of the National Academy of Sciences},
  110(19):7586--7591, 2013.

\end{thebibliography}

\end{document}